\documentstyle[12pt]{article}
\begin{document}
\newcommand{\newc}{\newcommand}
\newc{\zeus}{{\sc Zeus}}
\newc{\gev}{\,GeV}
\newc{\rp}{$R_p$}
\newc{\rpv}{$\not\!\!R_p$}
\newc{\rpvm}{{\not\!\! R_p}}
\newc{\rpvs}{{\not R_p}}
\newc{\ra}{\rightarrow}
\newc{\lsim}{\buildrel{<}\over{\sim}}
\newc{\gsim}{\buildrel{>}\over{\sim}}
\newc{\esim}{\buildrel{\sim}\over{-}}
\newc{\lam}{\lambda}
\newc{\lsp}{{\tilde\Lambda}}
\newc{\chio}{{\tilde\chi_0}}
\newc{\oc}{{\cal {O}}}
\newc{\msq}{m_{\tilde q}}
\newc{\mpl}{M_{Pl}}
\newc{\mw}{M_W}
\newc{\pho}{{\tilde\gamma}}
\newc{\dbar}{{\bar d}}
\newc{\ubar}{{\bar u}}
\newc{\half}{\frac{1}{2}}
\newc{\third}{\frac{1}{3}}
\newc{\quarter}{\frac{1}{4}}
\newc{\beq}{\begin{equation}}
\newc{\eeq}{\end{equation}}
\newc{\barr}{\begin{eqnarray}}
\newc{\earr}{\end{eqnarray}}
\newc{\ptmis}{\not\!\!{p_T}}
\newc{\mc}{\multicolumn}
\newcommand{\nnrpv}{${R}_p \hspace{-.9em}/$ \hspace{.5em}}
\newcommand{\nrpv}{$R_p$}
\newcommand{\photino}{{\tilde\gamma}}
\newcommand{\selectron}{{\tilde e}^-}
\newcommand{\squark}{\tilde q}
\newcommand{\sneutrino}{\tilde \nu}
\newcommand{\neutralino}{{\tilde \chi}^{0}_{i}}
\newcommand{\chargino}{{\tilde \chi}^{\pm}_{i}}
\newcommand{\pchargino}{{\tilde \chi}^{+}_{i}}
\newcommand{\nchargino}{{\tilde \chi}^{-}_{i}}


\title{Signals for Supersymmetry at HERA}
\author{Herbert Dreiner$\dag$ and Peter Morawitz$\ddag$}
\date{{\small $\dag$ Theoretische Physik, ETH-H\"onggerberg, CH-8093 Z\"urich\\
$\ddag$ Department of Physics, University of Oxford,\\
  1 Keble Rd, Oxford OX1 3RH}}
\maketitle

\begin{abstract}
We consider the baryon parity signals at HERA for the case of the MSSM
production mechanisms and the decays via the lepton number violating
couplings $L_iQ{\bar D}$. We can probe
very small Yukawa couplings $\lam' \gsim 3\cdot 10^{-6}$, limited only by
the decay length of the LSP. We assume the LSP to be the lightest
neutralino and study its decays in detail. We present the matrix
element squared for the tree-level decay amplitude of a generally mixed
neutralino explicitly. We find that the branching fraction to
charged leptons strongly depends on the SUSY parameters and can differ
significantly
from  the naively expected $50\%$.  The SUSY mass reaches of the studied
processes
 in the  ZEUS detector at HERA  were found to be:
$(m({\tilde e}, {\tilde \nu})+m({\tilde q}))\leq  170\gev$, $195\gev$ and
$205\gev$
 for the $L_\tau Q{\bar D}, L_\mu Q{\bar D}$  and $L_e Q{\bar D}$ couplings
respectively.
These  are well above existing limits on R-parity violating (\rpv) SUSY from
previous
experiments.   We conclude that HERA offers a {\it very promising} discovery
potential for \rpv\   SUSY.
\end{abstract}

\newpage

\section {Introduction}
When extending the Standard Model to include supersymmetry \cite{susy} the
superpotential
contains terms  which lead to unsuppressed proton decay. The most elegant and
economic
solution to this problem is to impose a discrete symmetry which disallows
 the dangerous $\Delta B \not=0$ terms. Several discrete symmetries have been
proposed,
with R-parity ($R_p$) \cite{susy} and Baryon Parity ($B_p$)
\cite{rp1,baryonparity} being the
most common. Imposing $R_p$ results in the minimal supersymmetric standard
model (MSSM); in the case of $B_p$ the superpotential contains the following
lepton number violating terms beyond those of the MSSM
\cite{gaugterms,rp1}
\beq
\lam_{ijk}' L_iQ_j{\bar D}_k,\quad \lam_{ijk} L_iL_j {\bar E}_k.
\label{eq:operators}
\eeq
$L,Q$ denote the left-handed lepton- and quark-doublets, respectively; ${\bar
E} ,{\bar D}$ denote the electron and down-like quark singlets respectively.
$\lam,\lam'$ are Yukawa couplings and $i,j,k=1,...,3$ are generation indices.
Theoretically it is still an unresolved question which is the preferred or
proper symmetry.\footnote{Experimentally clearly no statement can be made
since supersymmetry has not been found.} Presumably it will be
determined from the embedding of the supersymmetric standard model
in a more fundamental theory at a higher energy scale. In fact, recent work
in this direction hints at $B_p$ being the preferred symmetry \cite{ibanross}.

In light of this theoretical dilemma we propose to consider the full
phenomenological possibilities, in particular to include the $B_p$
signals in all supersymmetric searches. There is a growing amount of
work in this direction for searches at hadron colliders
\cite{rp3}-\cite{rplhc} and  at
$e^+e^-$-colliders \cite{rplola,tata} though clearly significantly less than
for
$R_p$.

In this paper we focus on promising $B_p$ signals at HERA. HERA is particularly
sensitive to the first operator in Eq.(\ref{eq:operators}) since it couples to
both leptons and quarks. When including a substantial coupling $\lam_{ijk}'$
the phenomenology changes in two main respects: (1) resonant squark production
is possible at substantial rates; (2) the lightest supersymmetric particle
(LSP)
is not stable and can decay in the detector.

Resonant squark production at HERA has been considered in
\cite{joanne,herastop}
and was studied in detail in \cite{butter,butter2}, where cascade
decays of the squarks to the LSP have been  included. The squark mass reach was
found to be \cite{butter2}
\beq
m_{\tilde q}\leq 270\gev \quad{\rm for}\quad \lam'\geq0.08,
\eeq
and the reach in the Yukawa coupling in this mode is
\beq
\lam'\geq 5.3\cdot10^{-3}\quad {\rm for}\quad m_{\tilde q}\simeq 100\gev.
\eeq

In this paper we fully exploit the second possibility, the decay of the
LSP. We consider the associate pair production of supersymmetric particles via
supersymmetric gauge couplings
\barr
e^-+q&\ra &{\tilde e}^- + {\tilde q}, \label{eq:ncprod} \\
e^-+q& \ra& {\tilde \nu} + {\tilde q}'. \label{eq:ccprod}
\earr
followed by the decay within the detector of the scalar fermions
(${\tilde f}={\tilde e},{\tilde\nu},{\tilde q}$) via the LSP to a
$R_p$-even final state.\footnote{The direct decay of the sfermions via
the operator $LQ{\bar D}$ is strongly suppressed for the small $\lam'_{ijk}$
we are considering. For a detailed discussion see \cite{rphadron,rplhc}.}
 For example in the case of a dominant operator $L_iQ_j{\bar D}_k$
\beq
{\tilde f}\ra f + {\tilde \chi}^0_1 \ra f + \{(e^\pm_i+u_j+{\bar d}_k) {\mbox{
or  }}
(\nu_i+d_j+{\bar d}_k) \}.
\eeq
We have denoted the LSP by ${\tilde\chi}^0_1$ the conventional notation for
the lightest neutralino. The main signal for these events will then be a high
$p_T$
charged lepton.

The mechanisms (\ref{eq:ncprod},\ref{eq:ccprod}) have been studied in detail
in the context of $R_p$, where the
dominant signals are missing $p_T$ \cite{chll,ruckl,heraws,montag}. These
studies have generally lead to the conclusion that there is little discovery
potential for
supersymmetry  at HERA beyond the limits set by the experiments at the
Tevatron. In \cite{butter,butter2} this was shown not to be the case if  $R_p$
is violated.
One essential point of this paper is to provide further viable signals at HERA
for the case of $B_p$.

In studying the $B_p$ signals we go beyond previous work
\cite{butter,butter2,dp}
in the following respects:
\begin{itemize}
\item The production cross sections  (\ref{eq:ncprod},\ref{eq:ccprod})  are
independent of the Yukawa
coupling. Thus it is possible to probe extremely small couplings,
which are only limited by the requirement that the LSP decay within
the detector, {\it i.e.} $\lam'\gsim 3\cdot 10^{-6}$. Recall, the smallest
Yukawa coupling in the SM, the electron-Higgs coupling, is just of this
order.

\item  Since the lepton number violating coupling is not involved in
the production cross section, we are able to explore the full generation
structure of the operator $L_iQ_j{\bar D}_k$. We must only require
$j\not=3$ since a top quark in the final state is beyond the kinematic
reach of HERA. In particular we can also study the $L_\tau Q_j{\bar D}_k$
operator, which  leads to a tau lepton in the final state.

\item We have considered a general neutralino as the LSP (as opposed to
restricting the study to a photino). This has two main effects in $B_p$
studies. (1) For the operator $LQ{\bar D}$ the branching fraction for the LSP
decay to charged leptons very strongly depends on the mixing parameters of
the neutralino sector.
(2) For a dominantly Zino/Higgsino LSP and small LSP masses
 the decay rate (proportional to a SM Yukawa
coupling squared) can become so small that the LSP does not decay within
the detector. This severely limits the range of couplings that can be
probed by the $B_p$ high $p_T$ charged lepton signals.

\item We have included the charged current production mechanism
(\ref{eq:ccprod}).
In $R_p$ this is not very promising since it is
difficult to distinguish from the SM charged current events. However, in
$B_p$ both the sneutrino and the squark decay via the LSP to predominantly
visible final states. This increases the overall event rate by about a factor
 2.5 \cite{bartl.cc}.
\end{itemize}

The clear disadvantage of employing the mechanisms (\ref{eq:ncprod}) and
(\ref{eq:ccprod}) is that the mass reach is severely
restricted by the kinematic limit of HERA, since two sfermions must be
produced.
This corresponds directly to the
problem encountered in the context of the MSSM. However, since we are
now considering the case where the LSP {\it decays} the previous MSSM squark
mass bounds derived mainly from searches for missing $p_T$ signals at CDF
must be reexamined. We discuss this in more detail below.

In our analysis we shall make the following simplifying assumptions
\begin{itemize}
\item[{\bf (A1)}]
We shall assume the LSP is a neutralino \footnote{This is a strong assumption,
since the LSP is not stable and the bounds on
relic densities of charged or coloured particles are no longer relevant.}. In
some regions of the MSSM
parameters the LSP is a chargino. We do not study these supersymmetry models.
\item[{\bf (A2)}] Of the 27 operators $L_iQ_j{\bar D}_k$ we shall assume in
turn that
one operator dominates \cite{rphadron}, and the others are negligible. We thus
have
18 separate scenarios to consider ($j\not =3$).
\item[{\bf (A3)}] The cascade decay of the sfermions to the LSP is
100\% of the branching fraction.
\end{itemize}

Our paper is outlined as follows. In Section 2 we summarise the relevant
previously existing bounds on the $B_p$ parameters. In Section 3 we
review the work on the production mechanisms
Eqs.(\ref{eq:ncprod},\ref{eq:ccprod}).
In Section 4 we present a detailed
analysis of the LSP decays for the operator $LQ{\bar D}$. We focus on the
branching fraction to charged leptons and on the decay length.
Combining the
last two sections we examine  the  signals, which we compare with possible
backgrounds
 in Section 5. In Section 6 we formulate  appropriate
cuts to extract the signals. This is the core of the analysis.
We  discuss the results  and conclude in
Section 7.

\section{Existing Bounds}
The existing bounds on the operators $\lam_{ijk}' L_iQ_j{\bar D}_k$
are of two kinds: (a) indirect bounds from low-energy processes involving
virtual supersymmetric particles; (b) direct bounds from collider
searches. Both sets of bounds make the simplifying assumption ({\bf {A2}}) of
a single dominant \rpv-operator.

(a) The most stringent relevant indirect bounds are \cite{vernon}
\barr
\lam_{L_eQ_i{\bar D}_j}' & \leq & \left\{
{\begin{array}{l}
0.06 \times \left[ \frac{{\tilde m}}{200\gev}\right],\quad i=1, \\
0.52 \times \left[ \frac{{\tilde m}}{200\gev}\right],\quad i=2,3
\end{array}} \right. \\
\lam_{L_\mu Q_i{\bar D}_j}' &\leq& \left\{
{\begin{array}{l}
0.18 \times \left[ \frac{{\tilde m}}{200\gev}\right],\quad i=1, \\
0.44 \times \left[ \frac{{\tilde m}}{200\gev}\right],\quad i=2,3.
\end{array}}
\right.
\earr
Note that the bounds scale linearly  in the relevant supersymmetric
scalar fermion masses ${\tilde m}$. And there are {\bf no} indirect
bounds on the tau-operators $L_\tau Q_i {\bar D}_j.$

(b) The most detailed collider bounds have been determined by D.P. Roy
\cite{dp}.
Here use is made of the CDF top-quark
search which sets bounds on any di-lepton production beyond that of the SM.
At the Tevatron, the MSSM pair production of squarks, followed by the
 \rpv\  cascade decay via the LSP would lead to a di-lepton signal. The
bound on the di-lepton production  directly  translates to a bound  on the
squark pair
production cross section, which to a good approximation only depends on the
squark masses. The bound on the squark masses thus obtained is
\beq
{\tilde m}_q \geq 100\gev,\quad
L_eQ_i{\bar D}_j ,L_\mu Q_i{\bar D}_j.
\eeq
And again there is no bound for the tau operators. In addition there
are model independent bounds from LEP, which hold for all operators
\cite{lepbound}
\beq
{\tilde m}_{e}\geq45\gev,\quad {\tilde m}_q \geq 45\gev.
\eeq
Thus our best bounds are
\beq
{\tilde m}_e+ {\tilde m}_q \geq  \left\{
{\begin{array}{l}
145\gev, \quad L_{e,\mu}Q{\bar D}, \\
90\gev,\quad \,\,\,\, L_{\tau}Q{\bar D}.
\end{array}}
\right.
\eeq

Besides studying SUSY bounds at HERA as opposed to the Tevatron, we
differ from \cite{dp} in the following two important points.
First, we consider a general neutralino LSP. This has a significant effect on
the branching fractions of the LSP in the case of the operator\footnote{That
the operator $LQ{\bar D}$ is special in this respect was already pointed out
in \cite{butter,butter2}.} $LQ{\bar D}$. It can also have a significant effect
on the
lifetime, and thereby the decay length of the LSP. The reach in coupling
constant in our analysis only depends on the decay length (see also  Section
4.3).
Second, we also consider the tau
operator $L_\tau Q{\bar D}$. HERA has an inherently cleaner environment than
the Tevatron and thus has a non-trivial discovery potential for $L_{\tau}Q{\bar
D}$.
Third, we consider {\it single} lepton signals and are thus less sensitive to
model-dependent variations in the  charged lepton branching fraction.

Further collider searches have been proposed for $e^+e^-$ \cite{tata} and
for hadron colliders \cite{rp3,me-and-dp}.
In particular \cite{tata} also does a detailed analysis
of the tau-number violating operators for LEP200. They consider gaugino pair
production which is complimentary to the dominant HERA signal.
The signal at LEP200 is insensitive to the squark mass and depends on the
slepton mass only via the propagator, however it strongly depends on the
(neutralino) LSP mass.

\section{Production Mechanism}
\label{prod.sec}

The dominant $R_p$ conserving SUSY production mechanisms at HERA are
the Neutralino Current (NC) and Chargino Current (CC) processes of
Figure \ref{nc.cc.fig}. Here $\tilde \nu$, ${\tilde e}_{L,R}$ and ${\tilde
q}_{L,R}$ are the sneutrino, the left and right handed selectron  and
squark respectively. In the MSSM there are two pairs of charginos,
$\chi^\pm_{1,2}$,
and four neutralinos, $\chi^0_{1...4}$. Following the notation of
 Haber and Kane \cite{susy},  masses and couplings depend on the
parameters $M'$ and $M$ (the U(1) and SU(2) gaugino parameters), the
Higgsino mass mixing parameter $\mu$, and $\tan(\beta)={\frac{v_1}{v_2}}$,
the ratio of the vacuum expectation values of the two Higgs doublets.
 We shall throughout assume the GUT relation \cite{GUT}
\beq
M' = {\frac {5} {3}} \tan^2(\theta_W) M,
\label{eq:GUT.relation}
\eeq
where $\theta_W$ is the electroweak mixing angle. Gaugino mixing has been
thoroughly
discussed in the literature, and we refer to \cite{susy,ruckl,Bartl.mix} for
details.

Processes (\ref{eq:ncprod}) and (\ref{eq:ccprod}) have been calculated by
various authors
\cite{Bartels,bartl.nc,bartl.cc}. As discussed in
\cite{bartl.workshop}, the cross sections exhibit a strong dependence
on the sum of the two final state SUSY particle masses, ($m_{\tilde
e}+m_{\tilde q}$) or ($m_{\tilde \nu}+m_{\tilde q}$). Furthermore they
show some dependence on the gaugino mass parameter $M'$, but are
relatively insensitive to the parameters $\mu$ and $\tan(\beta)$.
Figure \ref{sel.sneu.pair.xsec} shows a contour plot of the
SUSY cross section $\sigma_{e+q \ra {\tilde l} + {\tilde
q}'}$ (where ${\tilde l}$ can be a selectron or a sneutrino)
as a function of $M'$ and ($m_{{\tilde l}}+m_{{\tilde q}}$).
 The hatched region indicates where the mass of
the lightest neutralino  becomes
larger than $m_{{\tilde l}}$  and $m_{{\tilde q}}$,  the masses of the
scalars \footnote{Assuming  ${\tilde e}$, ${\tilde \nu}$ and  ${\tilde q}$ to
be degenerate in mass.}.
 Since we have assumed the lightest neutralino  to be the  LSP,
 we do not consider this region of parameter space.
 Note that within the region of interest to
HERA (scalar masses up to $\sim 200\gev$)
 cross section variations as a function of $M'$ are within a
factor of two for $M'>10\gev$. Figure
\ref{exp.no.evts.for.sel.sneus} shows the number of expected events
per 200$\,pb^{-1}$ (two nominal years of HERA running \footnote{Presently,
prior to  the 1994 running period, the integrated luminosity at HERA is
$1 pb^{-1}$.}) for a number of values $M'$ and their corresponding LSP masses.


The produced (on-shell) selectrons, sneutrinos and squarks subsequently
cascade-decay into lighter SUSY particles via  (making use of assumption
{\bf (A3)} in Section 1)
\begin{eqnarray}
{{\selectron_{L,R}}} & {\ra }& { e^- + \neutralino }
\label{eq:sel.neut1} \\
{{\sneutrino}} &  {\ra} & { \nu + \neutralino }
\label{eq:neut.neut1} \\
{\squark_{L,R} }& {\ra} &{ q + \neutralino } .
\label{eq:sq.neut1}
\end{eqnarray}
Eventually only one
type of SUSY particle remains, the LSP. In $R_p$ conserving models
the LSP escapes detection, and the resulting SUSY signals are
$e^-+1jet+ \mbox{ missing transverse momentum } ({\ptmis})$ for the
NC case, and $1jet+{\ptmis}$ for
the CC case. Several studies \cite{montag,bartl.workshop} have concluded that
HERA's eventual discovery reach on selectron, sneutrino and squark
production for R-parity conserving MSSM SUSY is $m_{\tilde e}+m_{\tilde q}<
200\gev$ and $m_{\tilde \nu}+m_{\tilde q}<180\gev$
at an integrated  luminosity of 200$\,pb^{-1}$.

In the following sections we shall concentrate on the decay of the LSP
via  \rpv\  couplings, including the full neutralino mixing of the MSSM.
We shall show that the squark mass discovery reach of the   LQD operator is
comparable to that  of the R-parity conserving case, for a very wide range
of Yukawa couplings.

\section{LSP Decays}
\label{Lsp.decay.sec}
\subsection{Matrix Element}
Here we present the matrix element squared for the decay of the LSP via the
operator $L_iQ_j{\bar D}_k$ to the charged lepton final state. We
consider a general neutralino LSP and we retain all final state fermion masses.
Thus we also include the ``higgsino''-like couplings which are
proportional to the fermion masses.
\begin{eqnarray}
 |{\cal M}(\chio\ra e_i u_j {\bar d}_k)|^2&=8c_fg^2\lam_{ijk}'^2 \left\{
\right.&  \\
 & \hspace{-2.5cm} D({\tilde u}_j)^2  e_i\cdot d_k&\hspace{-1cm}
 [(a(u_j)^2 + b(u)^2 )    \chio\cdot u_j
                 +   2 a(u_j) b(u) m_{u_j} M_{\chio}   \, ] \nonumber \\
& \mbox{}\hspace{-2.5cm} + D({\tilde d}_k)^2 e_i\cdot u_j & \hspace{-1cm}
[ (a(d_k)^2 + b({\bar d})^2)   \chio\cdot d_k-
                    2 a(d_k)b({\bar d}) m_{d_k}  M_{\chio}\, ]\nonumber \\
&  \mbox{}\hspace{-2.5cm} + D({\tilde e}_j)^2\hfil  u_j\cdot d_k
&\hspace{-1cm} [(a(e_i)^2 + b(e)^2)
          \chio\cdot e_i   + 2 a(e_i)b(e)  m_{e_i}  M_{\chio}\, ] \nonumber \\
 &\mbox{}\hspace{-2.5cm} -D({\tilde e}_i)D({\tilde u}_j) &
\hspace{-1cm}[ a(u_j)a(e_i)
m_{e_i}m_{u_j}  \chio\cdot d_k
          + a(u_j)  b(e)  m_{u_j}  M_{\chio} e_i\cdot d_k  \nonumber \\
  && \hspace{-1cm}       + a(e_i)  b(u)  m_{e_i}  M_{\chio}  u_j\cdot d_k
          + b(e)  b(u) \, g(u_j,\chio,e_i,d_k) ] \nonumber \\
 &\mbox{}\hspace{-2.5cm}  -D({\tilde u}_j) D({\tilde d}_k)&
\hspace{-1cm} [a(u_j) a(d_k) m_{u_j}m_{d_k}  \chio\cdot e_i
 - a(u_j)  b({\bar d})   m_{u_j}  M_{\chio}  e_i\cdot d_k  \nonumber \\
 && \hspace{-1cm} + a(d_k)  b(u)  m_{d_k}  M_{\chio}  e_i\cdot u_j
          - b(u)  b({\bar d}) \, g(u_j,\chio,d_k,e_i)]  \nonumber \\
 &\mbox{}  \hspace{-2.5cm}    -D({\tilde e}_i)D({\tilde d}_k)&
\hspace{-1cm} [-a(e_i)b({\bar d})
  m_{e_i} M_{\chio}  u_j\cdot d_k + a(e_i)  a(d_k)  m_{e_i}m_{d_k}
\chio\cdot u_j   \nonumber \\
 && \hspace{-1cm}\left. + a(d_k)  b(e)  m_{d_k}  M_{\chio}  e_i\cdot u_j
          - b(e)  b({\bar d})  \, g(\chio,e_i,u_j,d_k) ]  \nonumber \right\}
\end{eqnarray}
Here $c_f=3$ is the colour factor \footnote{A simpler form of this result
was obtained in Ref\cite{butter2} for the case of a photino LSP and neglecting
the final state fermion masses. The colour factor was omitted there by
mistake. This error does not affect any of the results presented there
since only branching fractions were employed, not the absolute decay rate.}
 and $g$ is the weak coupling constant. We have denoted
the 4-momenta of the initial and final state particles by their
particle symbols. The functions $D(p_i)$ denote the propagators squared
for particle $p$ and are given by
\beq
D({\tilde p}_i)^{-1}= M_\chio^2 + m_{p_i}^2-2\chio\cdot p_i - {\tilde
m}^2_{p_i}.
\eeq
The coupling constants $a(p_i),\,b(p)$ are for example given in \cite{gunhab}
and are listed in the Appendix for completeness.

The amplitude squared of the decay to the neutrino, $\chio\ra\nu_i d_j
\dbar_k$, can be obtained from
the above result by a set of simple transformations of the
4-momenta, the propagator functions $D$ and the couplings $a(p_i),\,b(p)$ :
$e_i\ra\nu_i$, $u_j\ra d_j$.
In order to determine the total decay rate these two modes must be
integrated over phase space, added, and then multiplied by a factor of
two, since the LSP is a Majorana fermion and can decay to the
conjugate final states.

The result for the operators $L_iL_j{\bar E}_k$ is completely analogous,
except the colour factor $c_f=1$. The result for the operators
${\bar U}{\bar D}{\bar D}$ is given in the Appendix.


\subsection{Charged Lepton Branching Fraction}
Before discussing the results of the calculation
in detail, we outline their general nature. We focus on the discussion
of the LSP decays to $\tau$'s via the $L_iQ_j{\bar D}_k$ operator ($j\not=3$).
Branching fractions to $e^\pm$ and $\mu^\pm$ (via the operators $L_eQ_j{\bar
D}_k$ and
$L_\mu Q_j{\bar D}_k$ respectively) are within a few percent of the tau
results. The only  noticeable difference enters in the Higgsino region,
where the LSP lifetime is  larger for decays to electrons
and muons than for decays to taus (by a factor $(m_{\tau}/m_{e,\mu})^2)$.

We have found that the
branching fraction of the LSP decay to $\tau$'s and $\nu_{\tau}$'s
strongly depends on the nature (i.e. the gaugino mixing parameters) of
the LSP. The LSP decay falls into one of the following three
distinctively different regions:
\begin{enumerate}
\item A region where the LSP predominantly decays to $\tau$'s.
Branching fractions of the LSP in this region are typically
 BF($\chi^0_1 \ra \tau^\pm + 2jets$)$\sim$70\%. Experimentally this
region is the most interesting one, because the event topology contains
an ``exotic'' lepton (the $\tau$).
\item A region where the LSP predominantly decays to $\nu_\tau$'s.
Branching fractions of the LSP in this region are typically
 BF($\chi^0_1 \ra \nu_\tau + 2jets$)$\sim$85\%.
One can still look for the signature of decays of the LSP to $\tau$'s, but the
bounds are significantly weaker.
\item A region in which Higgsino/Zino contributions to the LSP become
dominant and the LSP mass is small. In this region, the decay rate of the LSP
can become so
small that the LSP only decays within the detector for
anomalously
large  couplings $\lambda'$. Thus the dominant signature would again
be the missing  $p_T$  signal of the $R_p$ conserving MSSM scenario.
We will not consider this region in the
subsequent Monte Carlo (MC) study.
\end{enumerate}

 The neutralinos are
admixtures of their weak eigenstates ${\tilde \gamma}$, ${\tilde Z}$,
${\tilde H^0_1}$ and ${\tilde H^0_2}$. If the LSP dominantly consists
of the $\tilde \gamma$ component, branching fractions to $\tau^\pm$
are large, while a predominance of the ${\tilde Z}$ component makes
$\chi^0_1 \ra \nu_\tau + 2jets$ the more preferred decay channel. If the
LSP is more Higgsino-like, the scenario gets more complicated as the
branching fractions now also depend on both the vacuum expectation
values of the SM Higgs and the mass of the decay leptons.

We shall now discuss the results of the LSP calculation in detail.
The amount of mixing of ${\tilde \gamma}$, ${\tilde Z}$ and the two
Higgsinos
 is determined by the SUSY parameters $M'$, $M$, $\mu$ and $\tan(\beta)$.
Assuming the GUT relation (\ref{eq:GUT.relation}), we are left with
 the three  ``free'' parameters $M'$, $\mu$ and $\tan(\beta)$. Figure
\ref{lspdecay} shows the resulting LSP decay regions as a function of the SUSY
parameters for the dominant operator $L_{\tau}Q{\bar D}$.
The white region indicates where $m_{\chi^0_1} >
m_{\chi^\pm_1}$, the region where the mass of the lightest
neutralino is greater than the mass of the lightest chargino.
Since we have assumed that the LSP is a
neutralino, we do not consider this region in parameter space.
The branching fraction BF($\chi^0_1 \ra \tau^\pm + 2jets$) is less
than 30\% \footnote{We are interested in the pair production of
sparticles via processes (\ref{eq:ncprod}) and (\ref{eq:ccprod}), and
the  produced
selectrons, sneutrinos and squarks will subsequently decay to  pairs of
LSPs (plus other decay products). The condition that \mbox{BF($\chi^0_1 \ra
\tau^\pm +
2jets$)$<30\%$} corresponds to the condition that
\mbox{BF($2 \chi^0_1 \ra \mbox{1 or 2 }\tau^\pm\mbox{s } +
  jets $)$\lsim 51\%$}.}  in the light shaded area, and greater than 30\% in
the dark shaded regions. We have found that, over a wide range of
parameter space, the branching fractions change abruptly when the
nature of the LSP changes. This is demonstrated  in Figure \ref{lspdecay} by
solid
contours, where the ${\tilde \gamma}$, ${\tilde Z}$ and ${\tilde H}$
symbols on either side of the contours show which component's
contribution dominates the LSP. Furthermore we have  found that the
branching fractions are fairly constant in a given branching region,
and approximately BF($\chi^0_1 \ra \tau^\pm +2jets$)$\sim 70\%$ in the
dark regions, and BF($\chi^0_1 \ra \tau^\pm +2jets$)$\sim 15\%$ in the
light regions.

It is important to notice that these branching fractions significantly
modify the bounds obtained in Ref.\cite{dp}, where as a first approximation
a photino LSP was assumed.

\subsection{LSP Lifetime}
In order to get a qualitative understanding of the effects of the neutralino
mixing on the LSP lifetime let us consider first a photino LSP.
A photino LSP decays within the detector
(defined by the decay length $c\gamma\tau_{LSP}$ being less than $1\,m$,
$\tau_{LSP}$ is the LSP lifetime) provided \cite{rphadron,dawson}
\beq
\lam_{ijk}' \geq 3.7\gamma \cdot10^{-6}({\tilde m}_f/100\gev)^2
(45\gev/M_{LSP})^{5/2}
\eeq
where $\gamma$ is the Lorentz-boost factor in the laboratory, $c$ is the speed
of light. However, if the LSP is
a pure  Higgsino, then the decay rate is proportional to the
heaviest mass squared normalised to the Higgs vacuum expectation
value squared  $2m_i^2/v^2$, and the corresponding bound on $\lam'$ is
\beq
\lam_{ijk}' \geq 8.2\gamma \cdot 10^{-5}(m_\tau/(m_i))({\tilde m}_f/100\gev)^2
(45\gev/M_{LSP})^{5/2}.
\eeq
For the $L_\mu Q_1{\bar D}_1$ operator, the coupling $\lam'\gsim
1.4\cdot10^{-3}$,
which would still
allow a novel region to be explored. For the operator $L_eQ{\bar D}$ and a
Higgsino LSP, the single squark production followed by a direct \rpv\
decay should be more promising (discussed in \cite{herastop,joanne}).
 If the LSP does decay outside the detector we are  left either with the
leptoquark-like signature \cite{butter} or the standard missing $p_T$
signature of the MSSM.

In order to quantify the dependence on the MSSM mixing parameters,
we have plotted the regions of
parameter space (in black) in Figure \ref{lspdecay} for which the LSP
  (at a value of $\lambda'=0.03$) decays outside the detector, for
an operator $L_\tau Q{\bar D}$.
These regions
coincide with regions where the LSP is dominantly a mixture of
Higgsinos/Zinos and the LSP mass is small.  This is shown in Figure
\ref{lifetime}, which plots the LSP lifetime as a function of the Higgsino
mass parameter $\mu$. The two peaks at $\mu \sim 0\gev$ and $\mu \sim
100\gev$  correspond to the Higgsino/Zino region discussed above.
 The gap in the middle
at $\mu \sim 50\gev$ corresponds to the region where the LSP is not a
neutralino.
Figure \ref{lspdecay} also shows how far the black regions would
 extend (dashed lines) at
the very small values of the coupling constant  $\lambda'=3 \times 10^{-6}$.



\section{Signals and Backgrounds}
\label{sec.Signals.and.Backgrounds}
In order to investigate the full phenomenological consequences of the
LSP decay via the $L_iQ_j{\bar D}_k$ operator, we have used a $R_p$
violating SUSY generator to simulate the SUSY production mechanism of
Figures \ref{nc.cc.fig}a and \ref{nc.cc.fig}b and the subsequent \rpv\  LSP
decays
at the
parton level.
Throughout our analysis we have chosen to use the $MRSD$${^-}'$ structure
function set of reference \cite{mrsd} for the SUSY events and the
background samples.
We have generated the events at electron and proton beam energies of
$26.6 \gev$ and $820 \gev$ respectively.
 The SUSY events and potential backgrounds to the SUSY
signals were passed through a Monte Carlo detector simulation of the
ZEUS experiment based on the GEANT program \cite{geant}, and subsequently
reconstructed using the ZEUS offline programs. We will now proceed to
discuss the individual SUSY samples which we have used in our
analysis, their signatures, and the potential backgrounds to each of
the samples.

\subsection{SUSY Signals}
\label{SUSYsignals}
We have used the MSSM generator \cite{mssm.appendix}
 to simulate the \rpv\  SUSY
samples. In order to reduce the large SUSY parameter
space, we have made a number of simplifying assumptions in addition to
assumptions
{\bf  {(A1)-(A3)}}:
\begin{itemize}
\item
We have assumed sleptons (selectrons and sneutrinos) and the five
squarks to be degenerate in mass. Throughout we shall denote
the mass of the SUSY scalars as M$_{SUSY}=m_{\tilde l}=m_{\tilde q}$.
\item
We have only considered one set of gaugino mixing parameters ($M'=40\gev$,
 $\tan(\beta)=1$ and $\mu=-200\gev$) in the region where the LSP predominantly
decays to charged leptons, and one set of parameters ($M'=55\gev$,
 $\tan(\beta)$=4 and $\mu=+200\gev$) in the region
where the LSP predominantly decays to neutrinos. As discussed in
sections \ref{prod.sec} and \ref{Lsp.decay.sec}, variations  of the
gaugino mixing parameters within a given LSP decay region have a
relatively small
effect on the cross section and LSP branching fractions (we have found
that the combined changes in the cross section and the branching
fraction are within  a factor of two). This is to be  compared to the
large changes in the
cross sections as a function of  ($m_{{\tilde l}}+m_{{\tilde
q}}$)
which are approximately three  orders of magnitudes over
the SUSY mass scale of interest to HERA ({\it cf.} Fig. 2).
\end{itemize}

We now turn to the signatures of the SUSY events.
{}From the production mechanisms of Figure \ref{nc.cc.fig}a and
\ref{nc.cc.fig}b
two sparticles, a slepton and a squark, are produced. The sleptons and squarks
subsequently cascade-decay via processes
(\ref{eq:sel.neut1})-(\ref{eq:sq.neut1})
to neutralinos. Let us consider two specific examples of events in which
the LSP decays via the $L_\tau Q{\bar D}$ operator (see also Figure
\ref{hidious.decay.chain}):
\begin{eqnarray}
e^- + q \ra \selectron + \squark \ra  e^- + q + 2\neutralino & \ra & e^-
+ q + \tau + \nu_\tau + 4jets.
\label{eq:example1} \\
e^- + q \ra \selectron + \squark \ra  e^- + q + 2\neutralino & \ra &  e^-
+ q + 2\nu_\tau + 4jets.
\label{eq:example2}
\end{eqnarray}


Processes (\ref{eq:example1}) and (\ref{eq:example2}) are both
 (NC) processes.
Note that the pair of sfermions will eventually cascade-decay to {\it
two} LSPs.
Process (\ref{eq:example1}) has a
 tau in its final state. This
can subsequently decay to a positron or a muon via
\begin{eqnarray}
\tau^\pm & \ra & e^+ \nu_e{\bar \nu}_\tau  \vspace{1.0cm} (\mbox{BF} \sim
8.9\%)
\label{tau.to.pos} \\
\tau^\pm & \ra &  \mu^\pm {\nu}_\mu { \bar  \nu}_\tau \vspace{1.0cm} (\mbox{BF}
\sim
17.8\%)
\label{tau.to.mu}
\end{eqnarray}
and the branching fractions of the decays are shown in brackets. Our
overall signals for LSP decays via the $L_\tau Q{\bar D}$ are thus a
positron or a muon and  jet activity in the detector.
For the $L_e Q{\bar D}$ and the $L_\mu Q{\bar D}$ operator, our
signals are one or more positrons {\it or} one or more muons respectively
and   jet activity in the detector\footnote{Due to the large neutral
current deep-inelastic scattering background, the decay of the $\tau^-$
to an electron is considered a very difficult signature to observe
experimentally.}.

Having discussed the signals of interest, we now turn to the generated
SUSY samples.
We have generated  samples of SUSY events for each of the three  operators
above.
The $L_{e(\mu,\tau)}Q{\bar D}$ sample
 only contains events in which at least one of the two LSPs
has decayed to a positron (muon or tau respectively).

 For
each type of operator, $L_e Q{\bar D}$, $L_\mu Q{\bar D}$ and $L_\tau
Q{\bar D}$, we have generated  MC samples at a number of
SUSY masses   M$_{SUSY}$ ($=m_{\tilde l}=m_{\tilde q}$).
 The samples were generated with the gaugino parameters $M'=40\gev$,
 $\tan(\beta)$=1 and $\mu=-200\gev$, which
correspond to a region where the LSP
predominantly decays to charged leptons. The diagonalisation of the
gaugino mass matrix produces a  LSP with a mass of M$_{LSP}=44\gev$.
Table \ref{t:susy.gen} summarises the SUSY samples, their
 cross sections and the  expected number of events
 for an integrated   luminosity of 200pb$^{-1}$.

{\small{
\begin{table}
\begin{center}
\begin{tabular}{c||cccc|cc|cc}
\nnrpv  & & SUSY & parameters & &  & &  & \\
Operator    & M$_{SUSY}$ & $M'$ & $\tan(\beta)$& $\mu$ &
 Branch & $\sigma_{TOT}$ & Ngen & Nexp \\
           & (GeV) & (GeV) & & (GeV) &  (\%) & (pb) & & \\
\hline
\hline

$L_e Q {{\bar D}}$  & 45  & 40 & 1 & -200 & 0.80  &6.72 & 940 & 1344 \\
$L_e Q {{\bar D}}$  & 65  & 40 & 1 & -200 & 0.80  &1.71 & 500 & 342 \\
$L_e Q {{\bar D}}$  & 80  & 40 & 1 & -200 & 0.80  &0.50 & 400 & 100 \\
$L_e Q {{\bar D}}$  & 95  & 40 & 1 & -200 & 0.80  &0.11 & 100 & 22 \\
$L_e Q {{\bar D}}$  & 105 & 40 & 1 & -200 & 0.80  &0.03 & 100 &  6 \\
\hline
$L_\mu Q {{\bar D}}$  & 45  & 40 & 1 & -200 & 0.80  &10.10 & 1000 & 2020 \\
$L_\mu Q {{\bar D}}$  & 65  & 40 & 1 & -200 & 0.80  & 2.57 & 600 & 514 \\
$L_\mu Q {{\bar D}}$  & 80  & 40 & 1 & -200 & 0.80  & 0.75 & 400 & 150 \\
$L_\mu Q {{\bar D}}$  & 95  & 40 & 1 & -200 & 0.80  & 0.16 & 400 & 32 \\
$L_\mu Q {{\bar D}}$  & 100 & 40 & 1 & -200 & 0.80  &  0.09 & 200 & 18 \\
\hline
$L_\tau Q {{\bar D}}$  & 45 & 40 & 1 & -200 & 0.79  &10.09 & 2000 & 2018\\
$L_\tau Q {{\bar D}}$  & 65 & 40 & 1 & -200 & 0.79  & 2.56 & 1000 & 512 \\
$L_\tau Q {{\bar D}}$  & 80 & 40 & 1 & -200 & 0.79  & 0.75 & 1000 & 150 \\
$L_\tau Q {{\bar D}}$  & 85 & 40 & 1 & -200 & 0.79  & 0.46 & 400 & 92 \\
$L_\tau Q {{\bar D}}$  & 90 & 40 & 1 & -200 & 0.79  & 0.28 & 200 & 56 \\

\end{tabular}
\end{center}
\caption{\label{t:susy.gen} Generated R$_p$ violating  SUSY samples.
 M$_{SUSY}$ is the mass of the scalars;  $M'$,
 $\tan(\beta)$, $\mu$  are the gaugino mixing parameters;   Branch
 is the branching ratio of the LSP to a charged lepton
 $BF(\neutralino \ra l^\pm + 2 jets)$;  $\sigma_{TOT}$ is the total
 cross section of the sample;  Ngen  is the number of events
generated; Nexp is the number of events expected for an integrated  luminosity
of 200pb$^{-1}$,
two  nominal years of HERA running.}

\end{table}}}

\subsection{Backgrounds}
\label{backgrounds}
The most promising signatures of the \rpv\  SUSY events are a positron
or muon in the final state. Backgrounds to such a signature come from
``physics backgrounds'', which produce at least one positron or muon
directly, or from ``fake backgrounds'', in which either detector effects
fake positron or muon signals, or from positrons or muons produced
during the fragmentation process from
decays of secondary particles.
The individual background samples are described below.

In anticipation of a cut in total transverse energy $E_T$ in the
following analysis, and in order to keep the number of events in our
background samples down to a manageable size, we have applied a
$Q^2>1000{\gev}^2$ cut to the deep inelastic scattering neutral current
sample, and a $Q^2>500{\gev}^2$ cut to all other
samples at the generator level. Kinematically this corresponds to
a cut in total $E_T$, $E_T >60\gev$,   and  $E_T >45\gev$  respectively (see
Figure
\ref{fixed.et}). However, because a fraction of the transverse energy in each
event will be missed in the detector and also carried off by
neutrinos and  hence will not be
measured, the total visible (or reconstructed) transverse energy
 will be below the kinematically generated $E_T$ of Figure
\ref{fixed.et}. Also  calorimeter energy resolution effects
 will broaden the measured $E_T$ spectrum, and hence smear  events in
$E_T$ and shift them to higher/lower values of $E_T$.
Even taking both effects  into
account, our generator cuts on $Q^2$ are well below our final cut of $E_T >
90\gev$ in the following analysis.

\subsubsection {Physics Backgrounds}
We have considered two types of physics backgrounds. The first one is
heavy quark production via high Q$^2$ photon-gluon fusion \cite{bbar}. The most
dangerous case of photon-gluon fusion is expected to be $b{\bar b}$
production. The  produced $B^0$ can decay semileptonically to a
positron or a muon. We have also considered $c{\bar c}$ backgrounds,
but note that these are generally less dangerous  as their $E_T$
distribution is lower. The HERWIG generator \cite{herwig} was used to
generate  $b{\bar b}$ and $c{\bar c}$ events. Table \ref{t:back.gen}
summarises all background samples, their cross sections and the  expected
number of events
per luminosity of 200pb$^{-1}$, nominally two   years of HERA
running.

{\small{
\begin{table}
\begin{center}
\begin{tabular}{c||c|c|c|cc}
Sample Name & Generator & Generated Q$^2$ range (GeV$^2$) &
$\sigma_{TOT} (pb)$& Ngen & Nexp \\
\hline
\hline
$b {\bar b}$ & HERWIG& 500$\ra$87248 &  4.506 & 3395 &   901.2 \\
$c {\bar c}$ & HERWIG& 500$\ra$87248 & 37.39 & 4105 &  7478 \\
W ($e^+$, $\mu$) & EPVEC & 500$\ra$87248 & 0.219       &  400 &    43.8  \\
NC &           LEPTO & 1000$\ra$87248 &230.0 & 7149 & 46000 \\
CC &           LEPTO & 500$\ra$87248 & 49.5 & 3541 &  9900\\

\end{tabular}
\end{center}
\caption{\label{t:back.gen} Generated  background  samples. $\sigma_{TOT}$ is
the total
 cross section of the sample;  Ngen is the number of events
generated;  Nexp is the number of events expected for an integrated  luminosity
of 200pb$^{-1}$.}
\end{table}}}

The second physics background considered is W production \cite{wpro} via the
process
\begin{equation}
\gamma^* + q \ra W + q',
\end{equation}
where the virtual photon is radiated from the electron, and the W can
decay to give a positron or a muon. Cross sections are small compared
to the photon-gluon fusion backgrounds, but the $p_T$ of the charged leptons
is in general large. The EPVEC generator of reference
\cite{wpro} was used to generate the W sample. The sample only contains
events in which the W decays to a positron or a muon.

\subsubsection{``Fake Backgrounds''}
Fake signals of positrons or muons in the ZEUS detector can be caused
by a number of effects, and these include
\begin{itemize}
\item
Misidentification of the curvature (and thus charge) of high $p_T$
electrons in the tracking detector.
\item
Hadrons depositing energy in the  calorimeter with  shower
profile characteristics similar to that of a positron
(misidentification of positrons in the calorimeter).
\item
Punch-through of high energetic charged hadrons through the
calorimeter will give rise to tracks in the muon chambers, and hence
can lead to misidentified muons.
\end{itemize}

Furthermore, decays of secondary particles produced during
fragmentation processes can lead to genuine positrons and muons.
In order to investigate all the effects giving rise to fake
backgrounds, we have generated deep inelastic neutral current (NC) and
charged current (CC) events using the LEPTO program \cite{lepto}. Table
\ref{t:back.gen} summarises these  samples.


\section{Analysis}
\label{sec.analysis}
\subsection{Positron and Muon Identification}
Positron and muon identification play a key role in the extraction of
the SUSY signals. The identification algorithms have to give a high
purity in order not to contaminate the search sample with too much
fake background. With this in mind, we have chosen the
methods below for positron/muon identification.

Positron identification is achieved by first of all finding
 electromagnetic clusters  in the  calorimeter, e.g.
by finding shower profiles characteristic of those of electrons.
We  require the
cluster to have an energy above $5\gev$.
Tracks in the central drift chamber (CTD)
are then extrapolated   to the face of
the electromagnetic calorimeter. If a track matches such an
electron cluster, and the curvature of the track is positive, then
the cluster is identified as a positron.

Muon identification is achieved by matching CTD tracks with track
segments in the muon chambers. Note
however that only high energetic muons can penetrate the
calorimeter  before they reach the muon chambers which implies an indirect
 muon momentum  cut of $p_\mu >3-4\gev$.

\subsection{Cuts}
Before we go on to discuss how the SUSY signals can be extracted from
the background, we explain the cuts used in the subsequent analysis:

\begin{itemize}
\item
{\bf ``$e^+$ found and $\mu$ found''}: positron and  muon identification have
been discussed in the previous section.

\item
{\bf ``30\ GeV$< (E-p_z) < 60$\ GeV''}: $E$ is the sum of the total energies
 of the final state particles measured in the calorimeter,
 and $p_z$ is the  sum of the z component of
 momentum \footnote{The z-axis is defined to point along the proton
direction.}  of   all particles measured in the calorimeter. The $(E-p_z)$
variable  is an important quantity which
characterises the event \cite{zeus.paper}: for e-p events where all
final state particles are measured, $(E-p_z) = 2E_e$, where $E_e$ is
the energy of the initial state  electron.   Undetected particles
 which are emitted down the forward beampipe give a negligible loss in
$(E-p_z)$, while
 for example photoproduction (i.e. $\gamma$-p scattering) processes
in which the scattered electron remains in the beampipe, predominantly
result in  low values of $(E-p_z)$. The $(E-p_z)$ cut
 is very efficient in rejecting  photoproduction  events and other
backgrounds to the deep inelastic scattering processes of interest.

In our study we have used the $(E-p_z)$ cut to mainly reject charged current
(CC) events and W events. In both types of background a considerable
fraction of the energy is carried off by neutrinos, and hence will not be
detected. As a result the $(E-p_z)$ distribution of such events is
shifted towards lower values.

\item
{\bf ``circularity $>$ 1''}: One can define a 2-dimensional-like
sphericity,
 circularity \cite{circularity}, which is defined in the
range $0<$circularity$<1$.
The circularity  (transverse sphericity) is defined by
\begin{equation}
\mbox{circularity} = 2 \lambda_2
\end{equation}
in terms of the smaller eigenvalue, $\lambda_2$, of the two
dimensional sphericity tensor
\begin{equation}
S_{\alpha \beta}= \sum_i p_{i \alpha} p_{i \beta} / \sum_i p{_i}^2
\end{equation}
where $\alpha$, $\beta$ are Cartesian components, $i$ runs over  the
number of particles (i.e. reconstructed calorimeter clusters) in the
event, and $p$ are the four-momenta of the particles.
 The circularity   reflects the isotropy of the event.
  A  circularity of unity corresponds to a completely isotropic  event.
A three jet  event, for example, with the three jets pointing in
directions of the azimuthal angle, $\phi$, $\phi=0^\circ$, $\phi=120^\circ$ and
$\phi=240^\circ$
would be a good example of an event with a high value of  circularity.
A two jet event,  with a back-to-back jet configuration has a
circularity close to  zero.
We shall use this cut to distinguish between the SUSY events and
backgrounds. As will be discussed  below, the  SUSY events  are
very  isotropic, and thus have a high value of circularity.

\item
{\bf ``$E_T >$ 90\ GeV''}: $E_T$   is the total transverse energy measured
in the calorimeter. We also include the measurement of the muon track
in the
$E_T$  if a muon was found.
\end{itemize}

\subsection{Signal Extraction}
We first examine the characteristics of the positrons and muons in the
events.
Figure \ref{ptdis}a and \ref{ptdis}b show the muon and positron $p_T$
spectra for the SUSY samples at  M$_{SUSY}=80\gev$, correctly
normalised with respect to each other. Note that the muon
and positron $p_T$ spectrum of the $L_\tau Q{\bar D}$ sample is much
softer than the $L_\mu Q{\bar D}$ and $L_e Q{\bar D}$ $p_T$ spectra.
This is because the $\tau$'s have to decay via processes (\ref{tau.to.pos}) and
(\ref{tau.to.mu}) to positrons and muons, and hence  the energy of the
$e^+$ or $\mu$ is shared with the associated  neutrinos.


Even though the $p_T$ spectrum of the
$L_\mu Q{\bar D}$ sample is slightly harder than the $p_T$ spectrum of the
  backgrounds of section \ref{backgrounds}, we conclude that
the positron or muon $p_T$ variable is not a good discriminator
between the SUSY signals and the background processes.

Next we consider the circularity of the events. The decay process
(\ref{eq:example1}) gives  a typical SUSY event  topology, consisting of 5
jets, a $\tau$ lepton, an electron,
 neutrinos and the proton remnant. Unfortunately not all of the jets
are well resolved  - most of the jets will generally be  very soft. We have
found
that jet finding
algorithms can wrongly recombine the softer jets with harder jets,
and the jet multiplicity of such events can  be incorrectly determined. We have
run jet-finders on the  MC samples, and have found that
the jet multiplicity {\it can} be used to discriminate between the SUSY
events and backgrounds. However, we have also found that the
circularity variable is vastly superior to  jet multiplicity. This
can be explained as follows:

\begin{itemize}
\item
The high number of jets  result in a more isotropic event structure
for  the SUSY events. This makes the events more circular than
their respective backgrounds.
\item
Soft jets can not be resolved by jet-finders, but they do
contribute to the circularity of the event.
\end{itemize}

 Figure \ref{normal.circ} shows the
circularity of all combined backgrounds  and the SUSY $L_\tau Q{\bar
D}$ samples with M$_{SUSY}=45\gev$ and M$_{SUSY}=80\gev$. We observe two
features:
\begin{itemize}
\item
The circularity of the background falls off much more rapidly
than the circularity of the SUSY events, therefore a cut in circularity (chosen
to be at 0.1) will be  a good discriminator between  the SUSY
signal and background.
\item
Higher SUSY scalar masses have the effect of making the events more
circular - as a result higher M$_{SUSY}$ samples will be more
efficient in passing the circularity  cut.
\end{itemize}



Figure \ref{etplot}a shows the $E_T$ distribution of the backgrounds with
and without a  circularity cut. The cut reduces the
background by a factor of 10 at $E_T=60\gev$, and by a  factor of 18 at
$E_T=140\gev$. A similar plot is
shown for the SUSY $L_\tau Q{\bar D}$  samples with SUSY masses
M$_{SUSY}=45\gev$ and M$_{SUSY}=80\gev$. Again note that for the higher mass
sample the efficiency of the number of events surviving the
circularity cut at large values of  $E_T$ is higher  than for the lower mass
sample.


After  requiring an $e^+$  or $\mu$ and applying $(E-p_z)$ and circularity cuts
to the MC background samples, we are left with (102.4 $\pm$ 20.9) events.
Figure
\ref{finalet}a shows their $E_T$ distribution. The right-hand side of the
$E_T$ tail is
dominated by $b{\bar b}$ and $c{\bar c}$ events.  Note that due to the
higher cross section the $c{\bar c}$ sample is predominant in the
 $E_T$ background distribution up to an $E_T$ of $80\gev$. The
background events with $E_T>80\gev$ are all $b{\bar b}$ events, which
generally have a much harder  $E_T$ spectrum. We have chosen our final
 cut at $E_T>90\gev$. This  includes  (0.53 $\pm$ 0.31) $b{\bar b}$ events.
Figure \ref{finalet}b shows for comparison the $E_T$ distributions of the
three SUSY samples $L_e Q{\bar D}$, $L_\mu Q{\bar D}$ and $L_\tau
Q{\bar D}$ at  high mass ranges.


\subsection{Discovery Reaches}
We now apply the  analysis described above to all  MC
samples.
The only difference in the cut sequence applied to the $L_e Q{\bar D}$,
the  $L_\mu Q{\bar D}$  and the $L_\tau Q{\bar D}$ samples is
the positron/muon selection. We demand
\begin{itemize}
\item
One or more found  positrons for the $L_e Q{\bar D}$ sample.
\item
One or more found  muons  for the $L_\mu Q{\bar D}$ sample.
\item
One or more found  positrons {\it or} muons  for the $L_\tau Q{\bar D}$ sample.
\end{itemize}

Tables \ref{t:lqd111}, \ref{t:lqd211}  and \ref{t:lqd311} show the effects of
the cuts on the SUSY samples and the backgrounds.
 Figure  \ref{effic.plots}a shows  the efficiencies of the samples after
 cuts.  Normalising the cross sections of table \ref{t:susy.gen} to an
integrated  luminosity of L=200pb$^{-1}$, and folding in the
efficiencies of  Figure  \ref{effic.plots}a we obtain
 Figure \ref{effic.plots}b. We then use
 Poisson  statistics to investigate the potential discovery limit
 of the $L_e Q {{\bar D}}$ operator    at the 90$\%$ confidence level,
 by requiring at least 2.3 signal events and assuming zero background.
 The above condition holds  over the SUSY mass range $90\gev<(m_{\tilde
l}+m_{\tilde q})<205\gev$ for the $L_e Q {{\bar D}}$ operator and the
chosen set of gaugino parameters ($M'=40\gev$,
 $\tan(\beta)$=1 and $\mu=-200\gev$).  The
lower mass point at $(m_{\tilde l}+m_{\tilde q})=90\gev$
corresponds to the current model-independent LEP bounds.
Using  the same argument for the $L_\mu Q {{\bar D}}$ and the $L_\tau Q {{\bar
D}}$
sample, but requiring at least  3.1 signal events and assuming 0.53 background
events we can  determine  the approximate 90$\%$CL discovery reaches for the
$L_\mu Q
{{\bar D}}$ and the $L_\tau Q {{\bar D}}$ operators.
 Note that the gaugino parameters of the above SUSY MC samples
were chosen  in a region where the
 LSP predominantly decays to charged leptons.

{\small{
\begin{table}
\begin{center}
\begin{tabular}{c|rrrrr}
 cut  &  $b {\bar b}$ & $c {\bar c}$ &  W
 &  NC &
CC      \\
\hline
\hline
none &                       3395 & 4105 & 400 & 7149 & 3541 \\
$e^+$ found &                   22 &   24 & 118 &   55 &  8   \\
$30\gev < (E-p_z) < 60\gev$ & 20 &   21 &
 27 & 54  & 2 \\
circularity  $>$ 0.1 & 4 & 7 & 9 & 5
& 0 \\
$E_T > 90\gev$ & 0 & 0 & 0 & 0 & 0 \\
\hline
\hline
Exp. no. of evts & 0 & 0 & 0 & 0 & 0 \\

\end{tabular}
\end{center}

\begin{center}
\begin{tabular}{c|rrrrr}
 cut  &  $L_e Q {{\bar D}}_{45}$ & $L_e Q {{\bar D}}_{65}$ & $L_e Q {{\bar
D}}_{80}$ & $L_e Q {{\bar
D}}_{95}$ & $L_e Q {{\bar D}}_{105}$ \\
\hline
\hline
none &  940 & 500 & 400 & 100 & 100 \\
$e^+$ found &                  423 & 230 & 177 & 38 & 44 \\
$30\gev < (E-p_z) < 60\gev$ & 385 & 201 & 151 & 35 & 36 \\
circularity  $>$ 0.1 & 255 & 174 & 136 & 33 & 33 \\
$E_T > 90\gev$ & 67 & 124 & 124 & 32 & 32 \\
\hline
\hline
Exp. no. of evts & 95  $\pm$ 10 & 84.5  $\pm$ 6.5 & 31.0  $\pm$ 2.3 &
7.0  $\pm$ 1.0 & 1.9  $\pm$ 0.3 \\

\end{tabular}
\end{center}
\caption{\label{t:lqd111} The cut-sequence applied to the MC samples.
 Here ``$L_e Q {{\bar D}}_{45}$'' refers to the  SUSY  sample with
M$_{SUSY}=45\gev$, and LSP decays via the $L_e Q {{\bar D}}$ operator.
 The last row shows the expected number of events, properly normalised, after
cuts
for an integrated luminosity of   200$\,pb^{-1}$.}
\end{table}}}

{\small{
\begin{table}
\begin{center}
\begin{tabular}{c|rrrrr}
 cut  &  $b {\bar b}$ & $c {\bar c}$ &  W
 &  NC &
CC      \\
\hline
\hline
none &    3395 & 4105 & 400 & 7149 & 3541 \\
$\mu$ found &     126 &  71  &  65 &
29 & 11 \\
$30\gev < (E-p_z) < 60\gev$ &   114 &  59 & 24 & 20
& 3 \\
circularity  $>$ 0.1 &   42 & 18 & 3 & 2 & 0 \\
$E_T > 90\gev$ &           2 & 0 & 0 & 0 & 0 \\
\hline
\hline
Exp. no. of evts & 0.53 $\pm$ 0.31& 0 & 0 & 0 & 0 \\
\end{tabular}
\end{center}

\begin{center}
\begin{tabular}{c|rrrrr}
 cut  &  $L_\mu Q {{\bar D}}_{45}$ & $L_\mu Q {{\bar D}}_{65}$ & $L_\mu Q
{{\bar D}}_{80}$ & $L_\mu Q
{{\bar D}}_{95}$ & $L_\mu Q {{\bar D}}_{100}$ \\
\hline
\hline
none &    1000 & 600 & 400 & 400 & 200 \\
$\mu$ found & 395 & 191 & 114 & 119 &  50 \\
$30\gev < (E-p_z) < 60\gev$ & 330 & 157 & 92 & 102 & 44 \\
circularity  $>$ 0.1 & 177 & 131 & 78 & 88 & 31\\
$E_T > 90\gev$ &        27 & 71 & 58 & 79 & 31 \\
\hline
\hline
Exp. no. of evts & 54 $\pm$  10 & 60.7 $\pm$  6.7 & 21.8 $\pm$  2.7 &
6.3 $\pm$  0.6 & 2.8 $\pm$  0.5 \\

\end{tabular}
\end{center}
\caption{\label{t:lqd211} The cut-sequence applied to the MC samples.
 Here ``$L_\mu Q {{\bar D}}_{45}$'' refers to the  SUSY  sample with
M$_{SUSY}=45\gev$, and LSP decays via the $L_\mu Q {{\bar D}}$ operator.
 }
\end{table}}}

{\small{
\begin{table}
\begin{center}
\begin{tabular}{c|rrrrr}
 cut  &  $b {\bar b}$ & $c {\bar c}$ &  W
 &  NC &
CC      \\
\hline
\hline
none &    3395 & 4105 & 400 & 7149 & 3541 \\
$e^+$ or $\mu$ found &   148 &  95 &  183 &
84 & 19 \\
$30\gev < (E-p_z) < 60\gev$ &    134 & 80 &  51 & 74
& 5 \\
circularity  $>$ 0.1 &     46 & 25 & 12 & 7 &
0 \\
$E_T > 90\gev$ &         2 & 0 & 0 & 0 & 0 \\
\hline
\hline
Exp. no. of evts &   0.53 $\pm$ 0.31&  0 & 0 & 0 & 0 \\
\end{tabular}
\end{center}

\begin{center}
\begin{tabular}{c|rrrrr}
 cut  &  $L_\tau Q {{\bar D}}_{45}$ & $L_\tau Q {{\bar D}}_{65}$ & $L_\tau Q
{{\bar D}}_{80}$ & $L_\tau Q
{{\bar D}}_{85}$ & $L_\tau Q {{\bar D}}_{90}$ \\
\hline
\hline
none &  2000 & 1000 & 1000 & 400 & 200 \\
$\mu$ found & 185 & 74 & 92 & 33 & 14 \\
$30\gev < (E-p_z) < 60\gev$ & 166 & 53 & 71 & 27 & 11 \\
circularity  $>$ 0.1 & 105 & 40 & 58 & 21 & 8 \\
$E_T > 90\gev$ &        17 &  26 & 44 & 17 & 8 \\
\hline
\hline
Exp. no. of evts & 17.2 $\pm$ 4.2 & 13.4 $\pm$  2.6 & 6.6 $\pm$  0.9 &
3.9 $\pm$  0.9 & 2.2 $\pm$  0.8 \\

\end{tabular}
\end{center}
\caption{\label{t:lqd311} The cut-sequence applied to the MC samples.
 Here ``$L_\tau Q {{\bar D}}_{45}$'' refers to the  SUSY  sample with
M$_{SUSY}=45\gev$, and LSP decays via the $L_\tau Q {{\bar D}}$ operator.}
\end{table}}}

We have also investigated a region in SUSY parameter space  where the LSP
predominantly decays to neutrinos.
 Table \ref{mod.xsec} lists the SUSY
cross sections and LSP branching ratios in this region.
 Because the  parameters $M'=55\gev$,
 $\tan(\beta)$=4 and $\mu=+200\gev$ produce a LSP with M$_{LSP}=44\gev$,
the same LSP mass as  the generated MC samples of table \ref{t:susy.gen}, we
can use the efficiencies of Figure \ref{effic.plots}a to obtain  the
expected number of events after cuts of Figure \ref{effic.plots}c.

{\small{
\begin{table}
\begin{center}
\begin{tabular}{c||cccc|cc}
\nnrpv &   &  SUSY & parameters & &  &  \\
Operator    & M$_{SUSY}$ & $M'$ & $\tan(\beta)$& $\mu$ &
 Branch & $\sigma_{TOT}$  \\
           & (GeV) & (GeV) & & (GeV) &  (\%) & (pb)  \\
\hline
\hline

$L_e Q {{\bar D}}$  & 45  & 55 & 4 & +200 & 0.16  &1.67 \\
$L_e Q {{\bar D}}$  & 65  & 55 & 4 & +200 & 0.16  &0.43 \\
$L_e Q {{\bar D}}$  & 80  & 55 & 4 & +200 & 0.16  &0.12 \\
$L_e Q {{\bar D}}$  & 95  & 55 & 4 & +200 & 0.16  &0.03 \\
$L_e Q {{\bar D}}$  & 105 & 55 & 4 & +200 & 0.16  &0.01 \\
\hline
$L_\mu Q {{\bar D}}$  & 45  & 55 & 4 & +200 & 0.16  &3.12 \\
$L_\mu Q {{\bar D}}$  & 65  & 55 & 4 & +200 & 0.16  & 0.81 \\
$L_\mu Q {{\bar D}}$  & 80  & 55 & 4 & +200 & 0.16  & 0.24 \\
$L_\mu Q {{\bar D}}$  & 95  & 55 & 4 & +200 & 0.16  & 0.05 \\
$L_\mu Q {{\bar D}}$  & 100 & 55 & 4 & +200 & 0.16  &  0.03 \\
\hline
$L_\tau Q {{\bar D}}$  & 45 & 55 & 4 & +200 & 0.16  & 3.12 \\
$L_\tau Q {{\bar D}}$  & 65 & 55 & 4 & +200 & 0.16  & 0.81 \\
$L_\tau Q {{\bar D}}$  & 80 & 55 & 4 & +200 & 0.16  & 0.24 \\
$L_\tau Q {{\bar D}}$  & 85 & 55 & 4 & +200 & 0.16  & 0.15 \\
$L_\tau Q {{\bar D}}$  & 90 & 55 & 4 & +200 & 0.16  & 0.09 \\

\end{tabular}
\end{center}
\caption{\label{mod.xsec} Generated R$_p$ violating  SUSY samples in a
region of gaugino parameter space where the LSP predominantly decays
to neutrinos. }
\end{table}}}


Using the criteria previously described we now summarise   the maximum
discovery
reaches for the three operators $L_e
Q {{\bar D}}$, $L_\mu Q {{\bar D}}$ and $L_\tau Q {{\bar D}}$ in  the
two regions of gaugino parameter space:
\begin{eqnarray}
m({\tilde e}, {\tilde \nu})+m({\tilde q}) & \lsim &  205\gev
\mbox{\vspace{0.5cm} for dominant $L_e Q {{\bar D}}$, LSP decays to $e^\pm$},
\label{limit.111.dom} \\
m({\tilde e}, {\tilde \nu})+m({\tilde q}) & \lsim &  185\gev
\mbox{\vspace{0.5cm} for dominant $L_e Q {{\bar D}}$,  LSP decays to
$\nu_{e^\pm}$},
\label{limit.111.nondom} \\
m({\tilde e}, {\tilde \nu})+m({\tilde q}) & \lsim &  195\gev
\mbox{\vspace{0.5cm} for dominant $L_\mu Q {{\bar D}}$,  LSP decays to
$\mu^\pm$},
\label{limit.211.dom} \\
m({\tilde e}, {\tilde \nu})+m({\tilde q}) & \lsim &  175\gev
\mbox{\vspace{0.5cm} for dominant $L_\mu Q {{\bar D}}$,  LSP decays to
$\nu_{\mu^\pm}$},
\label{limit.211.nondom} \\
m({\tilde e}, {\tilde \nu})+m({\tilde q}) & \lsim &  170\gev
\mbox{\vspace{0.5cm} for dominant $L_\tau Q {{\bar D}}$,  LSP decays to
$\tau^\pm$},
\label{limit.311.dom} \\
m({\tilde e}, {\tilde \nu})+m({\tilde q}) & \lsim &  140\gev
\mbox{\vspace{0.5cm} for dominant $L_\tau Q {{\bar D}}$,  LSP decays to
$\nu_{\tau^\pm}$}.
\label{limit.311.nondom}
\end{eqnarray}

Note that the discovery reaches
(\ref{limit.111.dom})-(\ref{limit.311.nondom}) only hold for the
assumptions we have made in section \ref{SUSYsignals}, and only for
 the  particular choice of the gaugino mixing parameters, namely
($M'=40\gev$,
$\tan(\beta)$=1 and $\mu=-200\gev$) for dominant  LSP decays to $l^\pm$,
and ($M'=55\gev$,
$\tan(\beta)$=4 and $\mu=+200\gev$) for dominant  LSP decays to
$\nu_{l^\pm}$.
 We have also investigated the effect of
smaller LSP masses  on the
detection efficiencies of the samples, but found that they are
generally small: for a LSP mass of  M$_{LSP}=20\gev$ the correction to
the efficiencies of Figure \ref{effic.plots}a (at M$_{LSP}=44\gev$)
were found to be approximately greater than 0.6.
 Furthermore,  because cross section variations as a function of $M'$,
$\tan(\beta)$ and $\mu$ are relatively small (see section \ref{prod.sec}) and
changes in the LSP branching fractions to charged leptons   are fairly constant
in a
given characteristic region of the LSP (see section
\ref{Lsp.decay.sec}) we conclude that the discovery reaches  of
(\ref{limit.111.dom})-(\ref{limit.311.nondom}) are approximately
valid over a wide range of gaugino mixing parameter space within a
given characteristic region of the LSP.

\section{Discussion and Conclusions}
\label{sec.results}
We have investigated decays of the LSP as a general neutralino via the
$L Q {{\bar D}}$ operator. We have found that the results
depend on the mixing parameters of the neutralinos. Three distinct
regions in the parameter space characterise the LSP decay: (1) A
region in which the photino component of the LSP is dominant. In this
region branching fractions  to charged leptons are high,  typically $\sim
70\%$. (2)
A region in which the zino component of the LSP is dominant.  In this
region branching fractions to charged leptons are low, and typically $\sim
15\%$.
(3) A region in which the LSP is dominantly a  Higgsino/Zino admixture
at low LSP masses. In this region
the decay rate  of the LSP can become so small that the LSP only
decays within the detector for anomalously
large  couplings $\lambda'$, in which case the
only distinguishing feature
of \rpv\  is the single sfermion production. In  all other cases the
LSP decays in the detector for \rpv\   Yukawa couplings as low as $10^{-3}$.
At coulings $\lam' = 3\cdot  10^{-6}$ the LSP decays inside the detector in
only half the mixing scenarios.

Based on the above results we have investigated the discovery reaches
of HERA on the $L Q {{\bar D}}$ operator using  a full MC simulation
of pair produced SUSY signals with subsequent \rpv\   decays of the LSP.
We have considered the regions (1) and (2) of Section 4.2 in
gaugino parameter space, and have found that \rpv\  SUSY can be
discovered at HERA up to  combined SUSY masses
$(m({\tilde e}, {\tilde \nu})+m({\tilde q}))$ of
\begin{itemize}
\item
$170\gev$, $195\gev$, $205\gev$ for $L_{\tau} Q {{\bar D}}$, $L_{\mu} Q {{\bar
D}}$,
$L_{e} Q {{\bar D}}$ and mixing scenario (1)
\item
$140\gev$, $175\gev$, $185\gev$ for $L_{\tau} Q {{\bar D}}$, $L_{\mu} Q {{\bar
D}}$, $L_{e} Q {{\bar D}}$ and mixing scenario (2)
\end{itemize}
after two years of   HERA running at nominal luminosity.
 The discovery reaches are
to be compared with existing limits of $90\gev$, $ 145\gev$, $145\gev$
$(L_\tau Q{\bar D}, L_\mu Q{\bar D}, L_e Q{\bar D})$.
The  limits will only hold for the investigated    LSP  mixing
scenarios and
for the set of assumptions we have
made on the SUSY model, namely that (a) the LSP is a neutralino, that
(b) only one of the  27 operators $L_iQ_j{\bar D}_k$ is dominant, that
 (c) sleptons and squarks are degenerate in mass, and that  (d)
 the branching fraction of sfermion cascade decays to the LSP is $100\%$.

Because  HERA
 is an   inherently cleaner environment than the Tevatron,   there
are clear advantages for the search of \rpv\  SUSY signals.
Firstly our analysis demands  only one charged lepton from one of the
LSP decays. Consequently future limits  from HERA will be less model
dependent, and thus complimentary to existing  limits from the Tevatron
\footnote{The analysis of reference \cite{dp}
assumes {\it dilepton} signals coming from the
\rpv\  decay of two LSPs. }.
 Secondly our analysis is sensitive to LSP decays to $\tau$'s.  HERA is the
first
collider experiment which can directly probe the $L_\tau Q{\bar D}$.
The discovery potential of $\tau$ signals at HERA nicely ties in with
existing model independent  limits  on sfermion masses from LEP and
proposed searches for $\tau$ signals at LEP 200.

 We conclude that
HERA offers a very promising  discovery potential for direct  searches
for  \rpv\    SUSY,
which is nearly twice as high as the existing indirect bounds on
 the $L_\tau Q{\bar D}$ operator.

\section*{Acknowledgements}
We thank Neville Harnew for reading the manuscript and many helpful
suggestions, and Grahame Blair for useful discussions  and for  sharing his
 expertise on  SUSY generators  with us.

\section*{ Appendix}
Here we complete the discussion on the LSP decay matrix element.
The relevant coupling constants are given in Table \ref{Couplings}. Here
$a(p_i)$
denotes the coupling to the Higgsino which is proportional to the mass,
and $b(p_i)$ is the coupling to the gauginos. $N(1),N(2)$ are the
LSP admixtures of the Bino and the neutral Wino respectively. $N(3),N(4)$
are the Higgsino admixtures.

\begin{table}
\begin{center}
\begin{tabular}{|c|c|c|}\hline
& $a(p_i)$&$b(p_i)$ \\ \hline
$e_i$ &$ \frac{N(3)}{2 M_W\cos\beta}m_{e_i}$&$ -\frac{1}{2} (N(2)
 + \tan\theta_W N(1))$\\ && \\
${\bar e}_i$ &$\frac{N(3)}{2 M_W \cos\beta}\,m_{e_i}$&$-\tan\theta_W
N(1)$ \\ && \\
$\nu_i$ &$0$&$\frac{1}{2}(N(2)- \tan\theta_W N(1))$ \\ && \\
$u_i$ & $\frac{N(4)}{2M_W\sin\beta} m_{u_j}$ &$ \frac{1}{2}(N(2)
 + 1./3.\tan\theta_W N(1))$ \\  && \\
$\ubar_i$ &$\frac{N(4)}{2M_W\sin\beta} m_{u_j}$&$\frac{2}{3}\tan\theta_W
N(1)$ \\  && \\
$d_i$ &$ \frac{N(3)}{2M_W\cos\beta}  m_{d_k}$&$
-\frac{1}{2}(N(2) - \frac{1}{3}\tan\theta_W  N(1))$ \\  && \\
$\dbar_i$ & $ \frac{N(3)}{2M_W\cos\beta}  m_{d_k} $ &$- \frac{1}{3} \tan
\theta_W N(1)$ \\ \hline
\end{tabular}
\end{center}
\caption{\label{Couplings} The Coupling constants $a(p_i)$ and $b(p)$ used in
the LSP decay calculation.}
\end{table}
The matrix element squared for decay via the baryon
number violating operator (which we do not make use of in this paper)
is given by.
\barr
 |{\cal M}(\chio\ra {\bar u}_i{\bar d}_j{\bar d}_k)|^2 &= 8c_f g^2\lam''^2
\left\{\right.& \\
   & \hspace{-2.5cm}     D({\tilde u}_i)^2 {\bar d}_j\cdot {\bar d}_k&
\hspace{-1cm}  [ ( b(u)^2 +
a(u_1)^2 ) \chio\cdot {\bar u}_i   -2  b(u) a(u_i) m_{u_i} M_\chio ]
\nonumber \\
   &   \hspace{-2.5cm} +D({\tilde d}_j)^2  \ubar_i\cdot\dbar_k&\hspace{-1cm}
[ (a(d_j)^2 + b(d)^2)
\chio\cdot\dbar_j   -2  b(d)  a(d_j) m_{d_j} M_\chio ]
\nonumber \\
   &\hspace{-2.5cm} +D({\tilde d}_k)^2 \ubar_i\cdot\dbar_j&\hspace{-1cm}[
( a(\dbar_k)^2
+ b(\dbar)^2 ) \chio\cdot\dbar_k -2  b(\dbar) a(\dbar_k) m_{d_k}M_\chio]
\nonumber \\
   &\hspace{-2.5cm} + D({\tilde u}_i)D({\tilde d}_k)&\hspace{-1cm}[
 a(\ubar_i)a(\dbar_j)
\,g(\ubar_i,\chio,\dbar_j,\dbar_k) + b(\ubar)  b(\dbar) m_{u_i}m_{d_k}
\chio\cdot\dbar_k \nonumber \\
&&\hspace{-1cm}   - b(\ubar)  a(\dbar_j) m_{u_i} M_\chio \dbar_j\cdot\dbar_k
  -b(\dbar)  a(\ubar_i) m_{d_j} M_\chio \ubar_i\cdot\dbar_k   ]
\nonumber \\
   &\hspace{-2.5cm}   + D({\tilde u}_i)  D({\tilde d}_k)&  \hspace{-1cm}
[ a(\ubar_i)a(\dbar_k)\,g(u_i,\chio,d_k,d_j)+ b(\ubar)  b(\dbar)
m_{u_i}m_{d_k} \chio\cdot\dbar_j    \nonumber \\
   && \hspace{-1cm}-b(\ubar)  a(\dbar_k)m_{u_i} M_\chio \dbar_j\cdot\dbar_k
      -b(\dbar) a(\ubar_i) m_{d_k} M_\chio \ubar_i\cdot\dbar_j  ]
\nonumber \\
   &\hspace{-2.5cm} + D({\tilde d}_j) D({\tilde d}_k)&\hspace{-1cm}
 [ a(\dbar_j)a(\dbar_k) \,g(d_j,\chio,d_k,u_i)
+ b(\dbar) b(\dbar) m_{d_j} m_{d_k}  \chio\cdot\ubar_i
\nonumber \\
   && \hspace{-1cm}-b(\dbar)  a(\dbar_k) m_{d_j} M_\chio
\ubar_i\cdot\dbar_k
    \left.     -b(\dbar)  a(\dbar_j) m_{d_k} M_\chio\ubar_i\cdot\dbar_j  ]
\right\} \nonumber
\earr
The notation is as in Section 4.1. The colour factor $c_f=6$.

\newpage

\begin{figure}
\caption{\label{nc.cc.fig} }
\end{figure}
\begin{figure}
\caption{\label{sel.sneu.pair.xsec} }
\end{figure}
\begin{figure}
\caption{\label{exp.no.evts.for.sel.sneus} }
\end{figure}
\begin{figure}
\caption{\label{lspdecay} }
\end{figure}
\begin{figure}
\caption{\label{lifetime} }
\end{figure}
\begin{figure}
\caption{\label{hidious.decay.chain} }
\end{figure}
\begin{figure}
\caption{\label{fixed.et} }
\end{figure}
\begin{figure}
\caption{\label{ptdis} }
\end{figure}
\begin{figure}
\caption{\label{normal.circ} }
\end{figure}
\begin{figure}
\caption{\label{etplot} }
\end{figure}
\begin{figure}
\caption{\label{finalet} }
\end{figure}
\begin{figure}
\caption{\label{effic.plots} }
\end{figure}

\end{document}